\documentclass[aip]{revtex4-1}
\usepackage{hyperref}
\usepackage[pdftex]{graphicx}
\usepackage{epstopdf}
\usepackage{ulem}
\usepackage{textcomp}

\draft 

\begin{document}


\title{Novel method to study strain effect of thin films using a piezoelectric-based device and a flexible metallic substrate} 

\author{Kazumasa\,Iida}
\email[Electronical address:\,]{iida@mp.pse.nagoya-u.ac.jp}
\affiliation{Department of Materials Physics, Nagoya University, Chikusa, Nagoya 464-8603, Japan}
\affiliation{Department of Crystalline Materials Science, Nagoya University, Chikusa, Nagoya 464-8603, Japan}
\author{Yuwa\,Sugimoto}
\affiliation{Department of Crystalline Materials Science, Nagoya University, Chikusa, Nagoya 464-8603, Japan}
\author{Takafumi\,Hatano}
\author{Takahiro\,Urata}
\affiliation{Department of Materials Physics, Nagoya University, Chikusa, Nagoya 464-8603, Japan}
\affiliation{Department of Crystalline Materials Science, Nagoya University, Chikusa, Nagoya 464-8603, Japan}
\author{Marco\,Langer}
\author{Bernhard\,Holzapfel}
\author{Jens\,H\"{a}nisch}
\affiliation{Institute for Technical Physics, Karlsruhe Institute of Technology, Hermann-von-Helmholtz-Platz 1, 76344 Eggenstein-Leopoldshafen, Germany}
\author{Hiroshi\,Ikuta}
\affiliation{Department of Materials Physics, Nagoya University, Chikusa, Nagoya 464-8603, Japan}
\affiliation{Department of Crystalline Materials Science, Nagoya University, Chikusa, Nagoya 464-8603, Japan}


\date{\today}

\begin{abstract}
For applying tensile or compressive uniaxial strain to functional thin films, we propose a novel approach in combining a piezoelectric-based device and a technical metallic substrate used widely in the 2nd generation coated conductors (i.e. superconducting tapes). A strain-induced shift of the superconducting transition temperature of 0.1\,K for Co-doped BaFe$_2$As$_2$ was observed along [100] direction, corresponding to a uniaxial pressure derivative $dT_{\rm c}/dp_{100}=-4$\,K/GPa. For Mn$_3$CuN, a uniaxial strain derivative along [100] direction of the Curie temperature $dT_{\rm C}/d\epsilon_{100}=13$\,K/\% was observed. The current approach is applicable to various functional thin films in a wide range of temperatures.
\end{abstract}


\maketitle

Exploring the uniaxial pressure dependence of physical properties of functional materials gives an important clue for fundamental understanding of physics. To conduct such experiments, single crystals with reasonable volume are preferable. For some functional materials, however, the growth of sizable single crystals can be problematic, even though epitaxial thin films can be grown, e.g. Fe-based superconductors represented by NdFeAs(O,F),\cite{Kawaguchi-1, Kawaguchi-2} SmFeAs(O,F),\cite{Ueda} and antiperovskite manganese nitrides, Mn$_3$CuN.\cite{Hatano} Moreover, non-equilibrium phases, e.g. (Ba,$RE$)Fe$_2$As$_2$ ($RE$: rare earth elements = La, Ce, Pr and Nd)\cite{Katase, Katase-1} and FeSe$_{1-x}$Te$_x$ with low Te content\cite{Imai} are stabilized in the form of thin films. However, external pressure hardly transfers to strain in a thin film, since thin films are usually fabricated on substrates with high rigidity.

A common way to apply pressure to epitaxial films is to employ the lattice mismatch between substrate and film. But this method often (or strictly speaking always) requires the film thickness to be less than the critical value for which the coherent growth is maintained. Beyond the critical thickness, misfit dislocations or even cracks are introduced to partially relax the lattice stress, which may mask the interplay between the lattice strain and physical properties.\cite{Jan}

An alternative approach is to grow epitaxial thin films on piezoelectric substrates and to observe the changes in physical properties 
successively and reversibly by dynamic biaxial in-plane strain using the inverse piezoelectric effect. For instance, the Curie temperature of La$_{0.7}$Sr$_{0.3}$MnO$_3$ deposited on pseudocubic (001) Pb (Mg$_{1/3}$Nb$_{2/3}$)$_{0.72}$ Ti$_{0.28}$O$_3$ (PMN--PT) can vary in a range up to 19\,K.\cite{Thiele} Similarly, strain-induced shifts of the superconducting transition temperatures ($T_{\rm c}$) up to 0.4\,K for La$_{1.85}$Sr$_{0.15}$CuO$_4$ and 0.2\,K for BaFe$_{1.8}$Co$_{0.2}$As$_2$ were observed by applying in plane strain of 0.022 and 0.017\%, respectively.\cite{Trommler} However, the lattice parameter variation of the piezoelectric substrates is significantly reduced at cryogenic temperatures. For PMN--PT single crystals the strain $\epsilon_a$ for an electric field of $E=10$\,kV/cm reduced from 0.12\% at 300\,K to 0.022\% at 20\,K.\cite{Trommler} Here, the in-plane strain $\epsilon_a$ is defined by $(a_0-a_{\rm strained})/a_0$, where $a_0$ and $a_{\rm strained}$ are the in-plane lattice constant at $E=0$ and $E$$\neq$0, respectively.

\begin{figure}[b]
	\centering
		\includegraphics[width=15cm]{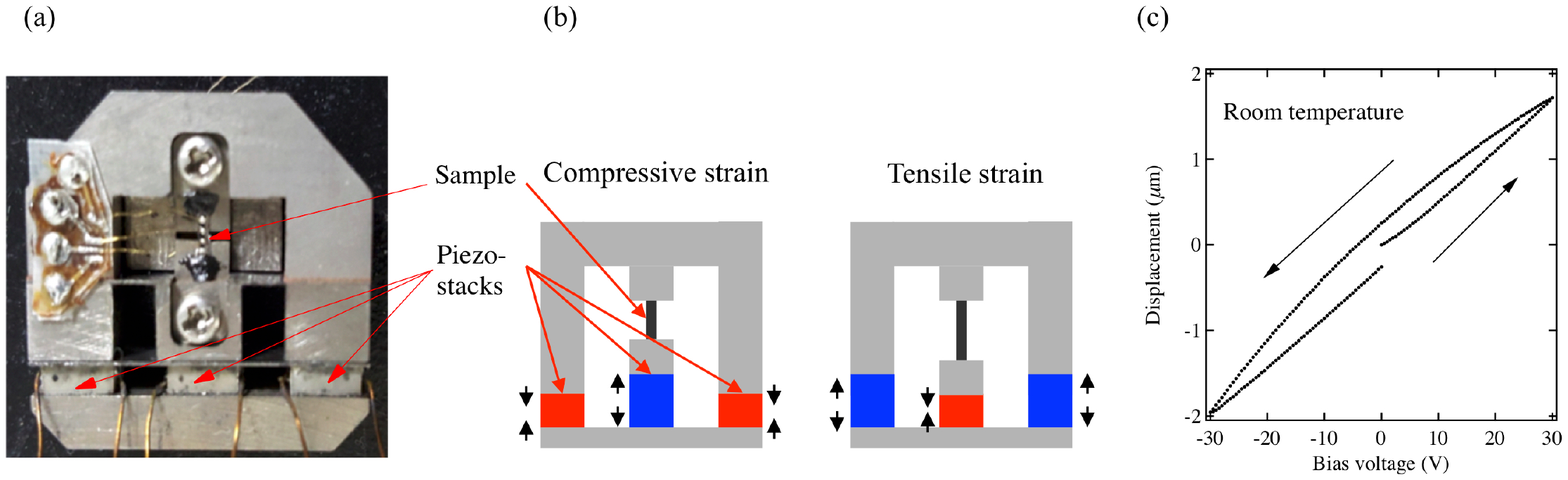}
		\caption{(Color online) (a) Top view of the piezoelectric-based apparatus for uniaxial strain experiments and (b) the corresponding schematic illustration. For applying a compressive (tensile) stress to the sample, the two outer piezostacks shrink (expand), whereas the central stack expands (shrinks) with bias voltage. (c) Bias voltage dependence of displacement for the small slab of IBAD-MgO (3.5\,mm long and 1\,mm wide) at room temperature.} 
\label{fig:figure1}
\end{figure}

To overcome those problems and also to investigate the effect of in-plane symmetry breaking on the physical properties, we report on a novel approach in combining a piezoelectric-based device\cite{Hicks} and a technical metallic substrate used widely in the 2nd generation coated conductors (i.e. superconducting tapes). The device has one central and two outer piezostacks as shown in fig.\,\ref{fig:figure1}(a). The central stack and the outer stacks generate compressive and tensile strains or vice-verse [fig.\,\ref{fig:figure1}(b)]. Since the piezostacks are longer than the effective sample length (see below), large stress can be applied to the sample even at low temperatures. Additionally, the technical substrate is flexible thanks to its low thickness of 100\,\textmu m, resulting in a high ability to transfer large amounts of strain to the film. Besides, MgO templates prepared on the technical substrate by ion beam assisted deposition (IBAD) are suitable for epitaxial growth of various materials. Hence our approach is applicable for a wide range of functional materials, whenever epitaxial thin films are realized on IBAD-MgO Hastelloy. 
Indeed, Kwon $et$ $al$. have fabricated Nd$_2$Fe$_{14}$B thin films on Mo-buffered IBAD-MgO Hastelloys and strained the substrate plastically ($\epsilon=2\%$) by an external load which significantly influenced the magnetic properties of the film.\cite{Kwon} Our approach extends this concept and allows a dynamic and reversible control of the in-plane uniaxial strain. In this study, an Fe-based superconductor, Co-doped BaFe$_2$As$_2$, and a magnetostrictive material, Mn$_3$CuN, have been chosen as test materials.

The uniaxial strain apparatus [fig.\,\ref{fig:figure1}(a)] composed of three piezostacks [Pb(Zr$_x$Ti$_{1-x}$)O$_3$, Piezomechanik GmbH] was constructed according to Ref.\,\onlinecite{Hicks}. A strain gauge was mounted on the backside of the device. The commercially available technical metallic substrates on which biaxially textured MgO templates of 160\,nm thickness were deposited by IBAD were provided by iBeam Materials, Inc. To avoid inter-diffusion between the metallic tape (Hastelloy C-276), and MgO the template contains Y$_2$O$_3$ layer of 1\,\textmu m thickness.\cite{Sheehan} The thermal expansion coefficient of Hastelloy, Y$_2$O$_3$, and MgO are 1.08$\sim$1.13$\times$10$^{-5}$\,/K,\cite{hastelloy} 0.81$\times$10$^{-5}$\,/K,\cite{Y2O3-1, Y2O3-2} and 1.24$\times$10$^{-5}$\,/K,\cite{Iida} respectively, indicating that the thermal expansion mismatches between Hastelloy and Y$_2$O$_3$, and between Y$_2$O$_3$ and MgO are nontrivial. However, the Y$_2$O$_3$ layer is polycrystalline and has a large thickness of 1\,\textmu m. Therefore, the strain due to  thermal expansion mismatch is negligible. 

The technical substrate was cut into a small slab of 3.5\,mm length and 1\,mm width by laser cutting, which was used as a test piece. The test piece was bridged across the gap between the two sample plates using Stycast 2850FT, as shown in fig.\,\ref{fig:figure1}(a). The gap between the sample plates, i.e. the effective length of the test piece, was around $L=2.0$\,mm. By applying a bias voltage of $\pm$30\,V at room temperature (RT), displacements of around 1.7\,\textmu m and --2.0\,\textmu m were observed, respectively (fig.\,\ref{fig:figure1}(c)), which corresponds to a maximum change in strain of $0.18$\%. As expected, the displacement was significantly reduced to $\pm$0.7\,\textmu m at $T=14$\,K even with applying a bias voltage of $\pm$110\,V, corresponding to a strain level of $1.4/2000\times 100=0.07$\%. However, this level of strain is large enough for observing shifts of $T_{\rm c}$ for Co-doped BaFe$_2$As$_2$.

\begin{figure}[b]
	\centering
		\includegraphics[width=7cm]{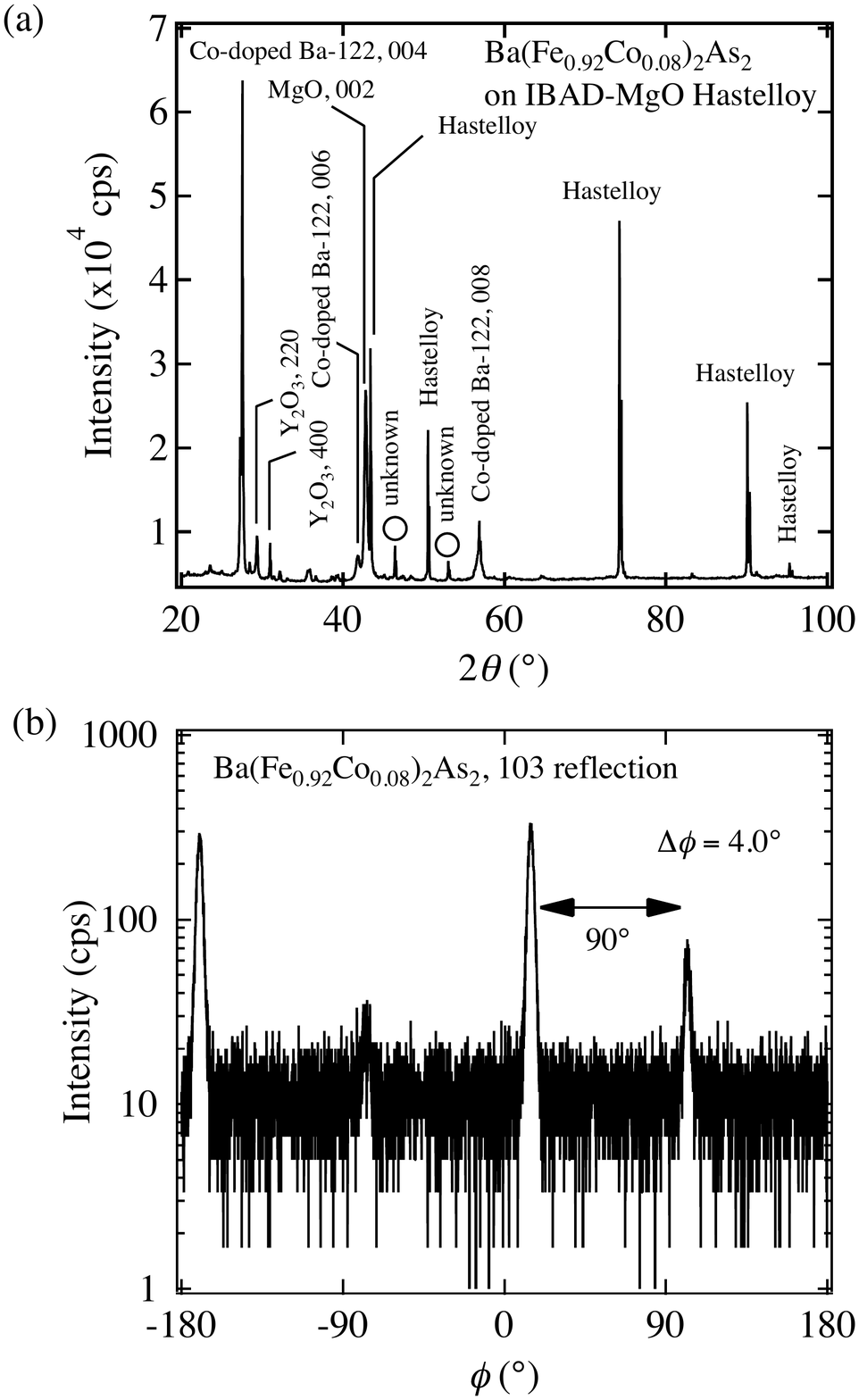}
		\caption{Summary of the structural characterization for Ba(Fe$_{0.92}$Co$_{0.08}$)$_2$As$_2$ fabricated on IBAD-MgO Hastelloy. (a) The $\theta$/2$\theta$-scan and (b)the azimuthal $\phi$-scan of the off-axis 103 reflection of Ba(Fe$_{0.92}$Co$_{0.08}$)$_2$As$_2$.} 
\label{fig:figure2}
\end{figure}

Co-doped BaFe$_2$As$_2$ thin films were deposited on IBAD-MgO Hastelloy at 800$^\circ$C by pulsed laser deposition (PLD) using a Nd:YAG 3rd harmonic laser ($\lambda = 355$\,nm) under ultra-high vacuum condition. This might be the first report ever on using a 3rd harmonic Nd:YAG for depositing BaFe$_2$As$_2$. The nominal composition of the sintered pellet used as PLD target was Ba(Fe$_{0.92}$Co$_{0.08}$)$_2$As$_2$. The energy density was around 4\,J/cm$^2$ at the surface of the target. Figure\,\ref{fig:figure2}(a) shows the $\theta/2\theta$--scans of the Co-doped BaFe$_2$As$_2$ (Ba-122) film on IBAD-MgO Hastelloy. All peaks except for a small amount of an unknown phase can be indexed with Co-doped Ba-122, MgO, Y$_2$O$_3$ and Hastelloy. The reflections from Co-doped Ba-122 were only 00$l$, indicative of the $c$-axis being oriented normal to the technical substrate. The azimuthal $\phi$-scan of the off-axis 103 reflection of Co-doped Ba-122 showed four peaks (fig.\,\ref{fig:figure2}(b)). These results prove that Co-doped BaFe$_2$As$_2$ was grown on IBAD-MgO Hastelloy with cube-on-cube configuration. The full width at half-maximum value ($\Delta\phi$) of the 103 reflection was around 4$^\circ$, which is similar to the underlying MgO. The films were cut into rectangular slices (5\,mm long and 1\,mm wide) by laser cutting. The longer direction is defined as being parallel to the crystallographic [100] direction.

\begin{figure}[b]
	\centering
		\includegraphics[width=7cm]{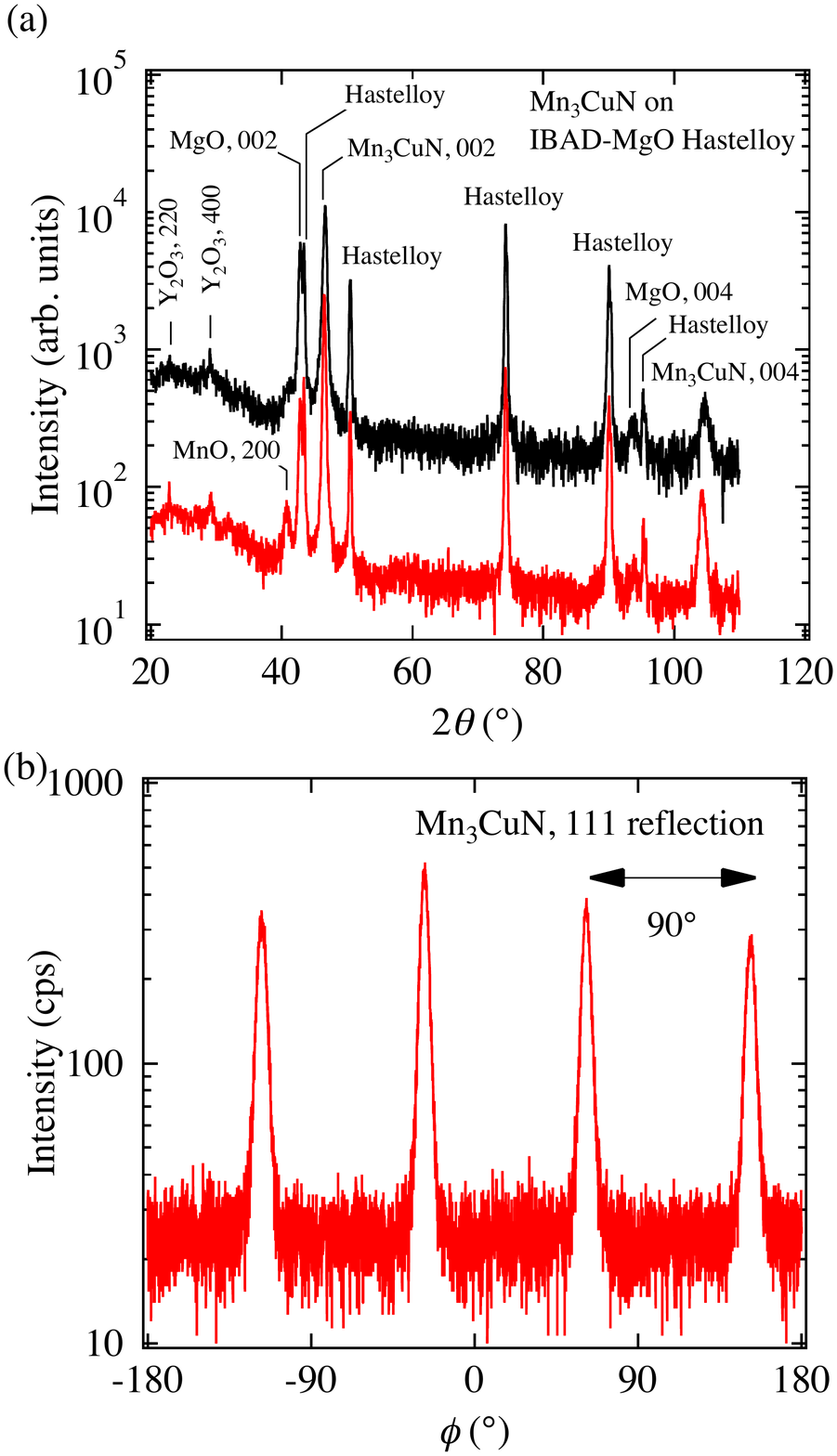}
		\caption{(Color online) Summary of the structural characterization for Mn$_3$CuN fabricated on IBAD-MgO template. (a) The $\theta$/2$\theta$-scan and (b)the azimuthal $\phi$-scan of the off-axis 111 reflection of Mn$_3$CuN.} 
\label{fig:figure3}
\end{figure}

Mn$_3$CuN thin films with 80\,nm thickness have been fabricated on IBAD-MgO templates at 200$^\circ$C in a custom-designed reactive sputtering chamber equipped with a high magnetic field of $\sim$1\,T at the surface of the target.\cite{Matsuda, Yanagi} Mn:Cu=4:1 alloy was employed as the target, and sputtering was conducted in Ar/N$_2$ gas mixture. The as-grown film was sealed together with Ti powder as oxygen absorber in a quartz tube, which was evacuated to a pressure level of 10$^{-4}$\,Pa. The whole arrangement was heated to 650$^\circ$C at a rate of 200$^\circ$C/min and subsequently furnace-cooled to room temperature. The process details can be found in Ref.\,\onlinecite{Hatano}. Structural characterization for the resultant film by XRD is summarized in fig.\,\ref{fig:figure3}. In fig.\,\ref{fig:figure3}(a), the 002 and 004 reflections of Mn$_3$CuN together with the 200 reflection of MnO were detected besides the peaks related to the technical substrate (i.e. Y$_2$O$_3$, MgO and Hastelloy). Although Ti powder was very effective for avoiding oxide formation in the post-annealing process, a small amount of MnO has formed. It is plausible that the oxygen came from the Y$_2$O$_3$ buffer layer. Nevertheless, the presence of MnO in our film does not affect the current investigation. Shown in fig.\,\ref{fig:figure3}(b) is $\phi$-scan of the 111 reflection for Mn$_3$CuN. A clear four-fold symmetry was observed. These results prove that Mn$_3$CuN can be grown epitaxially on IBAD-MgO Hastelloy. After structural characterization, the films were cut into rectangular slices in the same way as the Co-doped BaFe$_2$As$_2$ thin films.

\begin{figure}[t]
	\centering
		\includegraphics[width=7cm]{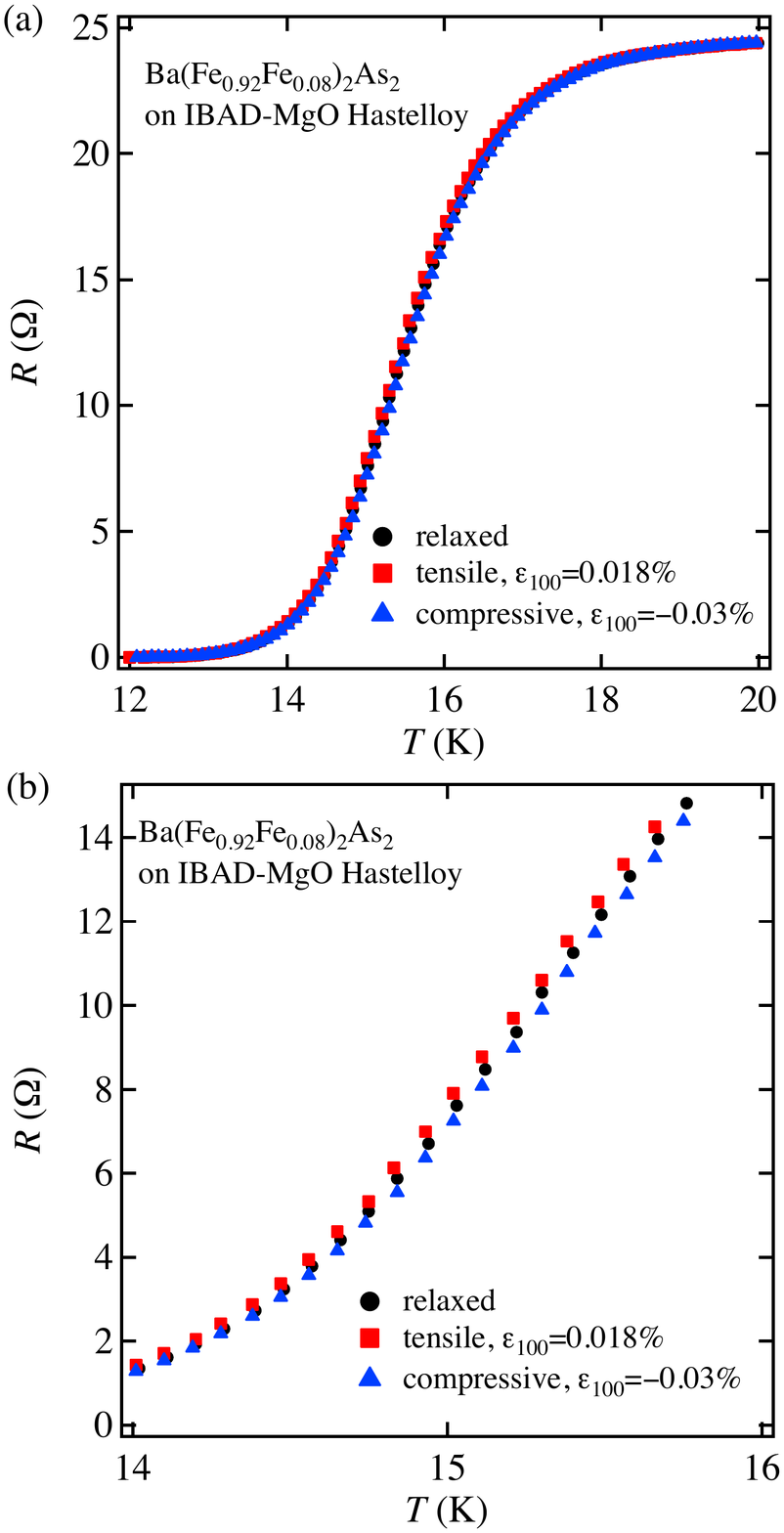}
		\caption{(Color online) (a) The resistance curves for Ba(Fe$_{0.92}$Co$_{0.08}$)$_2$As$_2$ thin films under various uniaxial strain state. (b) Enlarged view of the resistance curves in the vicinity of zero resistance.} 
\label{fig:figure4}
\end{figure}

Figure\,\ref{fig:figure4} shows the shift of the resistance curves for Ba(Fe$_{0.92}$Co$_{0.08}$)$_2$As$_2$ thin films on IBAD-MgO Hastelloy by uniaxial strain along [100] direction. The offset $T_{\rm c}$ defined as the intersection between the steepest slope of the resistivity curve and zero resistance were 14.2\,K for tensile ($\epsilon_{100}=0.018\%$) and 14.3\,K for compressive strain ($\epsilon_{100}=-0.03\%$), respectively. This tiny change is due to the small sensitivity of $T_{\rm c}$ against the strain along [100] direction. The strain-direction dependence of the $T_{\rm c}$ for P-doped Ba-122 single crystals showed that the change in $T_{\rm c}$ is the smallest for [100] direction compared to [110] and [001] directions.\cite{Kuo} A similar trend is expected for Co-doped Ba-122, although the doping element differs from P. On the assumption that the Young modulus ($E_{100}$) of Co-doped Ba-122 is 55\,GPa,\cite{Souza} the respective tensile and compressive stress values are calculated as 9.9$\times 10^{-3}$\,GPa and $-1.65\times 10^{-2}$\,GPa using Hooke's law. Accordingly, the uniaxial pressure derivative of the offset $T_{\rm c}$ approximates $dT_{\rm c}/dp_{100}=-4$\,K/GPa, which is in good agreement with the results for Ba(Fe$_{0.926}$Co$_{0.074}$)$_2$As$_2$ single crystals obtained from the specific heat and thermal expansion data using the Ehrenfest relation.\cite{Budko} Hence the current technique allows us to investigate the uniaxial strain dependence of $T_{\rm c}$.

\begin{figure}[t]
	\centering
		\includegraphics[width=7cm]{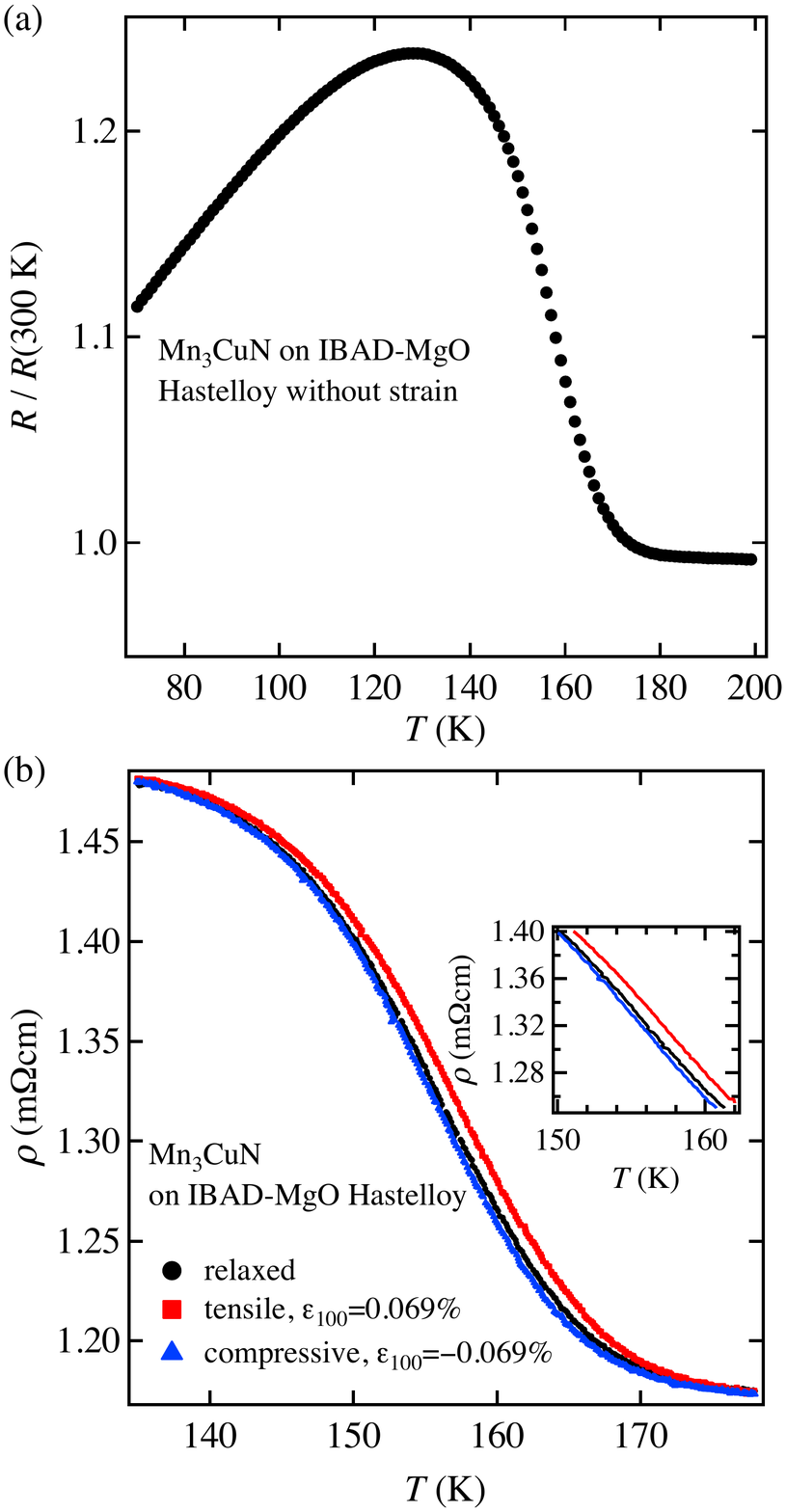}
		\caption{(Color online) (a) The normalized resistivity curve for the relaxed Mn$_3$CuN thin film. (b) The resistivity curves for the Mn$_3$CuN thin film under various uniaxial strain state. The inset shows an enlarged view of the resistivity curves in the temperature range $150\leq T \leq170$\,K.} 
\label{fig:figure5}
\end{figure}
 
The resistivity of the Mn$_3$CuN thin film suddenly increased at around 170\,K, which corresponds to the transition from high-temperature cubic to low-temperature tetragonal phase due to the Jahn-Teller effect [fig.\,\ref{fig:figure5}(a)].\cite{Hatano} This behavior is similar to the films on single crystalline substrates\cite{Hatano} and bulk polycrystalline samples.\cite{Chi, Asano} Although the temperature of the peak is almost unchanged by strain, a shift of the resistivity curves in the temperature range $140\leq T\leq170\,\rm K$ was observed as shown in fig.\,\ref{fig:figure5}(b): the tensile strain shifts the resistivity curve to higher temperature compared to the unstrained film. To the contrary, the compressive strain shifts the resistivity curve slightly to lower temperature. On the assumption that our Mn$_3$CuN films are not nitrogen-deficient, the structural transition temperature corresponds to the Curie temperature ($T_{\rm C}$). The evaluated $T_{\rm C}$ of the strained Mn$_3$CuN film were 167\,K (tensile strain, $\epsilon_{100}=0.069\,\%$) and 165.2\,K (compressive strain, $\epsilon_{100}=-0.069\,\%$). Here, $T_{\rm C}$ was defined as the intersection between the steepest slope of the resistivity curve and a constant resistivity of 1.18\,m$\Omega \cdot$cm (the resistivity value at 180\,K). Hence, the uniaxial strain derivative $dT_{\rm C}/d\epsilon_{100}$ along [100] without considering orthorhombic distortion is calculated as 13\,K/\%.

It is reported that $T_{\rm C}$ of epitaxial Mn$_3$CuN thin films grown on different substrates systematically changes due to biaxial strain.\cite{Hatano} The strain dependence of $T_{\rm C}$ without considering a Jahn-Teller strain is described by

\begin{equation}
T_{\rm C}=T_{\rm C}(0)+(2\epsilon_{100}+\epsilon_{001})\frac{\delta T_{\rm c}}{\delta \epsilon_{100}},
\end{equation}

\noindent
where $T_{\rm C}(0)=143$\,K is the Curie temperature of unstrained Mn$_3$CuN.\cite{Tsui}   
Using the reported values of $T_{\rm C}$ and the lattice parameters at RT for films on different substrates,\cite{Hatano}
the strain derivative is evaluated as $dT_{\rm C}/d\epsilon_{100}$=48\,K/\%. It is expected that this value does not change significantly near the transition temperature, since the linear thermal expansion coefficient of Mn$_3$CuN ($1.77\times 10^{-5}$\,/K)\cite{Chi} is comparable to that of MgO ($1.24\times 10^{-5}$\,/K).\cite{Iida} Nevertheless, the strain derivative derived from the study using single crystalline substrates is notably different from the results obtained from the uniaxial strain in this study. The reason for this discrepancy is not clear but might be due to nitrogen deficiency and/or off-stoichiometry. Further studies are necessary to elucidate the cause of this discrepancy.

In summary, we have proposed a novel approach combining a piezoelectric-based apparatus and flexible metallic substrates for investigating the in-plane uniaxial pressure dependence of the physical properties of functional thin films. The strain-induced shift of $T_{\rm c}$ along [100] direction was around 0.1\,K for Co-doped BaFe$_2$As$_2$, corresponding to a uniaxial pressure derivative $dT_{\rm c}/dp_{100}=-4$\,K/GPa. For the Mn$_3$CuN film, the uniaxial strain derivative $dT_{\rm C}/d\epsilon_{100}$ of the Curie temperature along [100] direction was 13\,K/\%. Our novel method demonstrates that uniaxial strain dependence of the physical properties of functional thin films can be investigated dynamically and reversibly, whenever epitaxial thin films are realized on IBAD-MgO Hastelloy. This approach thus greatly expands the experimental range of many functional thin films in allowing investigations on strain-driven changes in physical properties.

\begin{acknowledgments}
The authors would like to thank S. Aswartham and S. Wurmehl for preparing the PLD target. KI thanks Clifford Hicks for fruitful discussions. YS wishes to acknowledge the financial support of the Deutsche Akademische Austauschdienst (DAAD) for his stay at Karlsruhe Institute of Technology. This work was supported by a Grant-in-Aid for Challenging Exploratory Research Grant Number 15K13336 and for Scientific Research (B) Grant Number 16H04646 from the Japan Society for the Promotion of Science.
\end{acknowledgments}

\end{document}